\documentclass[twocolumn,showpacs,prl]{revtex4}
\usepackage[latin9]{inputenc}
\usepackage{color}
\usepackage{verbatim}
\usepackage{amsmath}
\usepackage{amssymb}
\usepackage{graphicx}
\usepackage{esint}
\usepackage{hyperref}

\makeatletter
\@ifundefined{textcolor}{}
{%
 \definecolor{BLACK}{gray}{0}
 \definecolor{WHITE}{gray}{1}
 \definecolor{RED}{rgb}{1,0,0}
 \definecolor{GREEN}{rgb}{0,1,0}
 \definecolor{BLUE}{rgb}{0,0,1}
 \definecolor{CYAN}{cmyk}{1,0,0,0}
 \definecolor{MAGENTA}{cmyk}{0,1,0,0}
 \definecolor{YELLOW}{cmyk}{0,0,1,0}
}

\usepackage{mathrsfs}

\draft

\makeatother

\begin{document}

\title{Manipulating Majorana Fermions in Quantum Nanowires with Broken Inversion
Symmetry}

\author{Xiong-Jun Liu and Alejandro M. Lobos}

\affiliation{Joint Quantum Institute and Condensed Matter Theory Center, Department
of Physics, University of Maryland, College Park, Maryland 20742, USA}

\begin{abstract}
We study a Majorana-carrying quantum wire, driven into a trivial phase
by breaking the spatial inversion
symmetry with a tilted external magnetic field. Interestingly,
we predict that a supercurrent applied in the proximate superconductor is able to restore
the topological phase and therefore the Majorana end-states. Using Abelian bosonization, we further confirm
this result in the presence of electron-electron interactions and
show a profound connection of this phenomenon to the physics of a  one-dimensional doped Mott-insulator.
The present results have important applications in e.g., realizing a supercurrent assisted braiding of Majorana fermions, which proves highly useful in topological quantum computation with realistic Majorana  networks.
\end{abstract}

\pacs{71.10.Pm, 74.45.+c, 74.78.Na, 03.67.Lx}


\maketitle

The study of topological superconductors (SCs) which host Majorana
zero bound states (MZBS) has developed into a remarkably lively and
rapidly growing branch of condensed matter physics, driven both by
the pursuit of exotic fundamental physics and the applications in
fault-tolerant topological quantum computation (TQC) \cite{Nayak,Ivanov,Sankar}.
MZBS exists in the vortex core of a two-dimensional (2D) $\left(p+ip\right)$-wave
SC \cite{Read}, and at the edges of a one-dimensional (1D) $p$-wave
SC \cite{Sengupta,Kitaev1}. However, intrinsic $p$-wave superconductivity
is not necessary to observe MZBS: recent proposals have shown the
equivalence of topological insulator/$s$-wave SC heterostructures
\cite{Fu1,Nilsson,Linder} and spin-orbit (SO) coupled semiconductor/$s$-wave
SC heterostructures with Zeeman splitting \cite{Sau0,Alicea,Roman0,Roman,Stanescu,Oreg,Potter}
to $p$-wave SCs. In such devices, the SO interaction drives the original
$s$-wave SC into an effective $p$-wave SC, leading to MZBS in the
case of odd number of subbands crossing the Fermi energy. It has been
predicted that an isolated MZBS in a topological SC can be
detected in differential tunneling conductance
$dI/dV$ at the interface with a normal contact, via the emergence
of a zero-bias peak (ZBP) of height $2e^{2}/h$ (at zero temperature)
\cite{Law,Flensberg,Wimmer,Liu,Lin}. Quite interestingly, recent experiments in semiconducting nanowires
(NWs)/$s$-wave SC heterostructures have shown a suggestive ZBP in the $dI/dV$
spectra \cite{Kouwenhoven,Deng,Das}, which disappears when the external magnetic field is tilted from the
direction of the NW and eventually aligned in the quantization axis of the SO coupling \cite{Kouwenhoven,Das}.

Motivated by these recent findings, in this work we investigate a Majorana-carrying
quantum NW driven into the trivial phase by a tilted
magnetic field which breaks 1D spatial inversion symmetry (SIS) \cite{Liu}, as observed in the experiment \cite{Kouwenhoven,Das}.
Interestingly, we show that a supercurrent applied in the SC
can compensate for the detrimental effects of the
tilted magnetic field, therefore restoring the MZBS.
Using Abelian bosonization we show the robustness of these findings in the presence of electron-electron (e-e) interaction, and provide insightful connections to the physics of doped 1D Mott insulators and the commensurate-incommensurate transition (CICT) \cite{giamarchi_book_1d}. We finally propose a supercurrent-assisted braiding (SAB) of MZBSs, which might have significant implications for TQC in realistic Majorana networks  \cite{Alicea1,Clarke}.

\begin{figure}[ht]
\includegraphics[width=1\columnwidth]{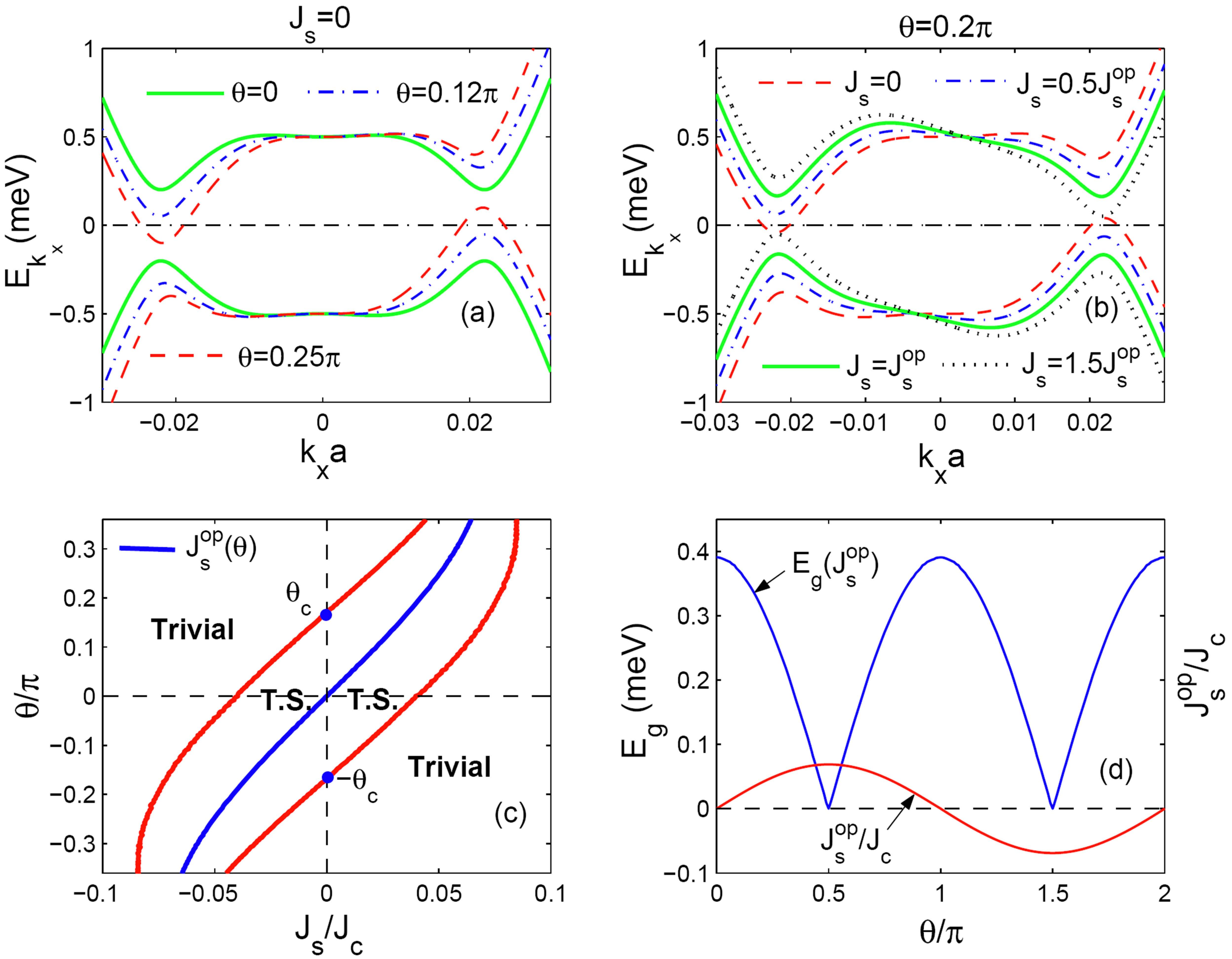} \caption{(a) Vanishing of the bulk gap by increasing tilt angle $\theta$;
(b) Restoring of the bulk gap at $\theta=0.2\pi$ by applying supercurrents;
(c) Phase boundary between topological superconducting (T.S.)
and trivial phases with a supercurrent; (d) Superconducting bulk gap
versus optimal supercurrent. Parameters are taken according to Ref. \cite{Kouwenhoven}: $V_{0}=1.0$meV,
$\Delta_{s}=0.5$meV, $\mu=0$, and SO energy $E_{{\rm so}}=m^{*}\lambda_{R}^{2}=0.1$meV
(a-d), resulting in a critical angle  $\theta_{c}\approx0.168\pi$ (cf. also \cite{note0}).}

\label{fig1}
\end{figure}

We start from the model of a 1D SO-coupled NW in proximity to an $s$-wave
SC, with a Zeeman field $\vec{V}=\left(V_{x},V_{y}\right)=V_{0}\left(\cos\theta,\sin\theta\right)$ given by an external magnetic
field tilted from the NW axis by an angle $\theta$. For $\theta=0$, a phase transition from a trivial
to a topological SC occurs by tuning $V_{0}$ beyond
a critical value $V_{c}=(\mu^{2}+|\Delta_{s}|^{2})^{1/2}$ \cite{Sau0,Alicea,Sato},
where $\mu$ and $\Delta_{s}$ are the chemical potential and induced
$s$-wave SC order parameter in the NW, respectively.
The Hamiltonian of the system is given by $H=H_{0}+H_{s}$,
where
\begin{equation}
  \begin{split}
H_{0}&=\int dx \mathbf{c}^{\dag}(x)\biggr[\frac{\partial_{x}^{2}}{2m^{*}}-\mu  +i\lambda_{R} \boldsymbol{\sigma}_{y}\partial_{x}+\vec{V}. \vec{\boldsymbol{\sigma}}\biggr]\mathbf{c}(x),\\
H_{s}&=\int dx\bigr[\Delta_{s}c_{\uparrow}(x)c_{\downarrow}(x)+\text{H.c.}\bigr],
\end{split}
\label{eq:H0}
\end{equation}
with $\mathbf{c}(x)=(c_{\uparrow}\left(x\right),c_{\downarrow}\left(x\right))$ the electron annihilation field operator, $m^{*}$ the effective mass of electrons in the
NW, $\lambda_{R}$ the Rashba SO coupling coefficient,
and $\vec{\boldsymbol{\sigma}}=(\boldsymbol{\sigma}_x,\boldsymbol{\sigma}_y,\boldsymbol{\sigma}_z)$ the vector of Pauli matrices. The term $V_{y}\sigma_{y}$, occurring
due to a finite tilt-angle $\theta$, breaks SIS of the NW \cite{Liu}. This can be seen directly
in $H_0$ under the 1D space-inversion transformation
$x\rightarrow-x$, $\left(y,z\right)\rightarrow\left(y,z\right)$,
which leads to $\left(k,\sigma_{y}\right)\rightarrow\left(-k,-\sigma_{y}\right)$
and $\sigma_{x}\rightarrow\sigma_{x}$, with $k$ the momentum along the NW. The broken SIS leads to an
asymmetric dispersion relation $\varepsilon_{k}^{\left(\pm\right)}\neq\varepsilon_{-k}^{\left(\pm\right)}$
for $H_{0}$, where $\varepsilon_{k}^{\left(\pm\right)}=k^{2}/2m^{*}\pm\sqrt{V_{x}^{2}+\left(V_{y}-\lambda_{R}k\right)^{2}}$.
Accordingly, the Bogoliubov quasiparticle spectra with a uniform
$\Delta_{s}$ are also asymmetric $E\left(k\right)\neq E\left(-k\right)$
[Fig.~\ref{fig1} (a)]. In particular, when $\theta$ is greater
than a critical value $\theta_{c}\left(V_{0},\Delta_{s},\mu\right)$,
the minimum (maximum) energy of the electron-like (hole-like) states
becomes negative (positive), and the bulk gap closes [red
dashed curves in Fig.~\ref{fig1} (a)]. This leads to a topological
phase transition at $\theta=\theta_{c}$, and for $\theta>\theta_{c}$
the system is a trivial SC. %
{}

We proceed to show that the topological phase can be restored at $\theta>\theta_{c}$
by a supercurrent $J_s$ applied in the proximate SC. In the presence of
a uniform $J_s$, the induced SC order parameter acquires a
position-dependent phase $\Delta_{s}\left(x\right)=|\Delta_{s}|e^{i\phi\left(x\right)}$,
related to the supercurrent through the relation $J_{s}=2n_{s}e\hbar\alpha\bigr[1-\left(\alpha\xi\right)^{2}\bigr]/m_{e}$
\cite{SC1}, with $\alpha=\nabla\phi\left(x\right)$ a uniform phase-gradient,
and $m_{e}$, $n_{s}$, and $\xi$ the electron mass, superconducting
carrier-density and coherence length in the bulk SC, respectively. The applied $J_s$ is required to be less than the superconducting critical current
$J_{c}=4n_{s}e\hbar/\left(3\sqrt{3}m_{e}\xi\right)$ \cite{SC1}.
The physics of the problem can be seen more transparently by projecting
$H$ onto the lower subband of the NW, $H\approx H^{\left(-\right)}=\sum_{k}\bigr[\varepsilon_{k}^{\left(-\right)}-\mu\bigr]c_{k,-}^{\dag}c_{k,-} +\frac{1}{2}\sum_{k}\bigr[\Delta_{s}e^{i\chi_{k}}c_{k,-}c_{-k-\alpha,-}+\text{H.c.}\bigr]$,
where $\chi_{k}=\tan^{-1}\left[\left(V_{y}-\lambda_{R}k\right)/V_{x}\right]$
and $\varepsilon_{k}^{\left(-\right)}>\varepsilon_{-k}^{\left(-\right)}$
for $k>0$ and $0<\theta<\pi$. %
For $J_s=0$, electron states with opposite momenta
$\pm k$ are off-resonant and the formation of Cooper-pairs with
zero center-of-mass momentum is weakened. For a supercurrent applied
along $+x$ direction (i.e., $\alpha>0$), the Hamiltonian
pairs up states with momenta $k$ and $-k-\alpha$ which are closer
in energy, favoring the formation of a Cooper pair with center-of-mass
momentum $\alpha$. A supercurrent therefore allows
to compensate for the band asymmetry induced by the tilted magnetic
field, strengthening the bulk gap in the NW.

In Fig.~\ref{fig1} (b) we show that the bulk gap, which vanishes for $\theta=0.2\pi$ at $J_{s}=0$, reopens in the presence of $J_s$ in the $+x$ direction, and attains its maximum at
the optimal value $J_{s}=J_{s}^{{\rm op}}$ (green solid line). Further
increasing $J_{s}$ suppresses again the bulk gap due
to an over compensation of the band asymmetry and induces again an off-resonant situation (black dotted line). Our results are summarized in Fig.~\ref{fig1}
(c), which shows the phase diagram of the NW as a function of
$J_{s}$ and $\theta$, with $\xi\leq10$nm the typical coherence length in NbTi SCs  \cite{McCambridge}. The blue curve
represents the optimal supercurrent $J_{s}^{{\rm op}}\left(\theta\right)$,
and the red curves give boundaries of the topological and trivial
phases. For $J_{s}=0$, the phase becomes trivial when $\theta_{c}\leq|\theta|<\pi/2$,
while applying a supercurrent along $+x$ (or $-x$, depending on
the sign of $\theta$) can restore the topological phase [Fig.~\ref{fig1}(c)]. In contrast, for $\theta=0$ the optimal supercurrent is $J_s^{\rm op}=0$, and applying a $J_s$ breaks the SIS and destabilizes the topological phase \cite{Romito}.
Fig.~\ref{fig1} (c) therefore provides a useful  guide to explore systematically the topological phase diagram in ongoing experiments \cite{Kouwenhoven, Deng, Das}.
The bulk gap $E_{g}\left(J_{s}^{{\rm op}}\right)$ versus  $J_{s}^{{\rm op}}$ is given in
Fig.~\ref{fig1} (d), from which one finds a vanishing $E_{g}\left(J_{s}^{{\rm op}}\right)$
only at $\theta=\pi/2,3\pi/2$, indicating that MZBSs can always
be restored by a supercurrent unless the magnetic field is perpendicular
to NW.

To determine if the above results are robust against  e-e interactions in the NW, we introduce here
the Abelian bosonization framework. At low energies, linearization
of the dispersion relation $\varepsilon_{k}^{\left(-\right)}$ around
the Fermi energy $E_{F}$ generates asymmetric left (right) Fermi
momenta $k_{L}\left(k_{R}\right)$ and Fermi velocities
$v_{L\left(R\right)}=\hbar^{-1} \partial_k\varepsilon_{k=k_{L\left(R\right)}}$ due to the broken SIS. We next introduce the standard
bosonic representation of left/right-moving fermions $c_{L/R}\sim\frac{1}{\sqrt{2\pi a}}e^{i\left(\mp\varphi-\vartheta\right)}$,
with bosonic fields $\varphi,\vartheta$
obeying the canonical commutation relation
$\left[\varphi\left(x\right),\vartheta\left(x'\right)\right]=i\pi\text{sign}\left(x'-x\right)/2$
and $a\sim k_{F}^{-1}$ the short-distance cutoff of the continuum
theory \cite{giamarchi_book_1d}. Physically, the field $\varphi\left(x\right)$
represents slowly-varying fluctuations in the electronic density $\delta\rho\left(x\right)=-\partial_{x}\varphi\left(x\right)/\pi$,
and $\vartheta\left(x\right)$ is related to the phase of the SC order
parameter through $c_{R}\left(x\right)c_{L}\left(x\right)\propto e^{i2\vartheta\left(x\right)}$.
With a short-range interaction $H_{\text{int}}=\pi U\int dx\; c_{R}^{\dagger}\left(x\right)c_{R}\left(x\right)c_{L}^{\dagger}\left(x\right)c_{L}\left(x\right)$
the low energy Hamiltonian is given in bosonic representation by \cite{Supplementary}
\begin{eqnarray}
H & = & \int dx\left[\frac{vK}{2\pi}\left(\partial_{x}\vartheta\right)^{2}+\frac{v}{2\pi K}\left(\partial_{x}\varphi\right)^{2}+\frac{\eta v}{\pi}\partial_{x}\varphi\partial_{x}\vartheta\right.\nonumber \\
 &  & \left.+\frac{\left|\Delta_{p}\right|}{\pi a}\sin\Big(2\vartheta\left(x\right)+\left(\alpha-\delta k_{F}\right)x\Big)\right],\label{eqn:H_bosonized}
\end{eqnarray}
where e-e interactions are encoded in the dimensionless Luttinger parameter
$K=\sqrt{(1-2U/v)/(1+2 U/v)}$.
$v=(\left|v_L\right|+\left|v_R\right|)/2$ is the average velocity, and $\Delta_{p}=\Delta_{s}\sin\bigr(\frac{\chi_{k_{L}}-\chi_{-k_{R}-\alpha}}{2}\bigr)$ is the effective $p$-wave SC order parameter.
The dimensionless parameter $\eta=(\left|v_L\right|-\left|v_R\right|)/(\left|v_L\right|+\left|v_R\right|)$
and $\delta k_{F}=k_{L}-k_{R}$ quantify the band-asymmetry .
When $\Delta_{p}=0$, the above model describes a Luttinger liquid (LL) fixed-point  with broken SIS and asymmetric dispersion relation, i.e., right- and left-going 1D plasmon excitations traveling at different
velocities \cite{Wen90,Blok90,Fernandez02_Asymmetric_LL}.
As shown in Ref.~\cite{Fernandez02_Asymmetric_LL}, the
asymmetric LL is a stable fixed-point with a well-defined Luttinger
parameter $K$ when $\eta^2+(2U/v)^2<1$. In general,
the SIS-breaking term $\sim\eta\partial_{x}\varphi\partial_{x}\theta$
tends to enhance the detrimental effects of the oscillatory factor $(\alpha-\delta k_F)x$ in Eq. \ref{eqn:H_bosonized} (see the Supplementary Material \onlinecite{Supplementary} for more details). However, for the typical parameters used in Fig.~\ref{fig1},
one can verify that $\eta<1\%$ at all tilt angles, and then $\eta\partial_{x}\varphi\partial_{x}\theta$ can be neglected in the following analysis.
We also note that for semiconductor NW, the system is generically far away from half-filling condition and the length of the wire $L \gg L_c \equiv \left|4\left(k_R+k_L \right)/2-2 \pi/a\right|^{-1}$, in which case the umklapp scattering  term $\cos\left[4\phi-2\left(k_{R}+k_L\right) x\right]$ becomes strongly oscillating at lengthscales larger than $L_c$ and averages out to zero~\cite{Gangadharaiah11_Majorana_fermions_in_1D_interacting_wires}.

For a small $\Delta_{p}$, the low-energy physics of
the model is captured by the perturbative renormalization-group (PRG)
approach around the LL fixed-point \cite{Gangadharaiah11_Majorana_fermions_in_1D_interacting_wires,Stoudenmire11_Interaction_effects_in_1D_wires_with_Majorana_fermions,Fidkowski11_Majorana_fermions_in_wires_without_LRO,Lobos12_Interplay_disorder_interaction_Majorana_wire}.
Implementing a standard PRG procedure that leaves
invariant the LL Gaussian fixed-point under the change in the short-distance
cutoff $a\left(\ell\right)=a_{0}e^{\ell}\rightarrow a\left(\ell+d\ell\right)$
allows to obtain the RG-flow equations: $dK/d\ell=y^{2}J_{0}\left(\delta pa\left(\ell\right)\right)$,
$dy/d\ell=\left(2-K^{-1}\right)y$ and $dv/d\ell=-y^{2}vKJ_{2}\left(\delta pa\left(\ell\right)\right)$,
with $\delta p\equiv\alpha-\delta k_{F}$ (see Ref.~\cite{Supplementary} for more details).
Here $J_{n}\left(z\right)$ is the $n$-th order Bessel function of
the first kind and $y\equiv\Delta_{p}a_{0}/v$ is a dimensionless
perturbative parameter which becomes relevant (in the RG sense) for
$K>1/2$ and $\alpha=\delta k_{F}$ \cite{Gangadharaiah11_Majorana_fermions_in_1D_interacting_wires,Stoudenmire11_Interaction_effects_in_1D_wires_with_Majorana_fermions,Fidkowski11_Majorana_fermions_in_wires_without_LRO,Lobos12_Interplay_disorder_interaction_Majorana_wire}.
Interestingly, our RG equations are analogous to those describing
the CICT in doped 1D Mott-insulating
systems after the rescaling $\tilde{K}=4K, \tilde{\vartheta}=\vartheta /2, \tilde{\varphi}= 2\varphi$, and the subsequent duality transformation $\tilde{\vartheta}\leftrightarrow \tilde{\varphi}, \tilde{K}\leftrightarrow 1/\tilde{K}$ \cite{japaridze_cic_transition,Pokrovsky1979,Schulz1980}.
The crucial term $\delta px$ in Eq. (\ref{eqn:H_bosonized}) plays
the role of the particle-doping (relative to half-filling case) in
the CICT, which has the effect of closing the Mott insulating gap.
Analogously, in our case a finite $\delta p$ may close the SC gap.
The condition $\alpha=\delta k_{F}$ (i.e. $\delta p=0$) determines
the optimal supercurrent
$\frac{J_{s}^{{\rm op}}}{J_{c}}  =  \frac{3\sqrt{3}}{2}\left[1-\left(\xi\delta k_{F}\right)^{2}\right]\xi\delta k_{F}$  in the bosonization approach,
for which the Majorana-carrying topological phase is maximally
restored. This result is independent of interactions, and relies
on the linearization of $\varepsilon_{k}^{\left(-\right)}$ around
$E_{F}$ (non-linearities may slightly correct the value of
$J_{s}^{{\rm op}}$).

We now estimate the critical value $\delta p_{c}$ for the topological phase transition.
At very small $\delta pa\left(\ell\right)\ll1$, the $\sin$ function
in Eq. (\ref{eqn:H_bosonized}) is weakly oscillating and the term
$\delta px$ can be dropped, rendering the RG equations similar to
the those of the (undoped) sine-Gordon model \cite{Gangadharaiah11_Majorana_fermions_in_1D_interacting_wires,Stoudenmire11_Interaction_effects_in_1D_wires_with_Majorana_fermions,Fidkowski11_Majorana_fermions_in_wires_without_LRO,Lobos12_Interplay_disorder_interaction_Majorana_wire}.
In that case and for $K>1/2$, we reach the strong-coupling regime $y\left(\ell^{*}\right)\sim1$ at the scale $\ell^{*}=\left(2-K^{-1}\right)^{-1}\ln\left(\xi_{\rm nw}/a_{0}\right)$ with $\xi_{\rm nw}=v/|\Delta_p|$, where the SC term $\Delta_{p}\sin2\vartheta$ dominates in Eq. (\ref{eqn:H_bosonized}).
In this regime, the value of $\vartheta\left(x\right)$ is pinned
to the classical minima $\vartheta\left(x\right)=\left\{ -\pi/4,3\pi/4\right\} $
of the $\sin2\vartheta$ potential, reflecting the underlying $\mathbb{Z}_{2}$
symmetry of the Majorana chain in the limit $L\rightarrow\infty$ \cite{Kitaev1,Fidkowski11_Majorana_fermions_in_wires_without_LRO,Lobos12_Interplay_disorder_interaction_Majorana_wire}.
As $\delta p$ increases, the regime $\delta pa\left(\ell\right)>1$
is eventually reached and the $\sin$ function becomes strongly oscillating
and averages to zero. At that point the above RG equations are no
longer valid and the renormalization of $y\left(\ell\right)$ must
be stopped \cite{giamarchi_book_1d}. The critical value $\delta p_{c}$
can be estimated from the condition $\delta p_{c}a\left(\ell^{*}\right)=1$, which implies that
\begin{eqnarray}
\delta p_{c} & \sim & \frac{1}{a_{0}}\left(\frac{a_{0}}{\xi_{\rm nw}}\right)^{\nu},\ \ \nu=\frac{1}{2-K^{-1}}.\label{eqn:scaling1}
\end{eqnarray}
This is an important result in our work. In particular, the noninteracting
case $U=0$ (or $K=1$) results in $\delta p_{c}\propto\xi_{\rm nw}^{-1}\propto\Delta_{p}$
, which has been confirmed by direct numerical calculation in the
noninteracting model. In the case $K\neq1$, and for fixed $y_{0}=y\left(\ell=0\right)$,
we observe that repulsive (attractive) e-e interaction destabilizes
(stabilizes) the topological phase, inducing a smaller (larger) $\delta p_{c}$.
Importantly, for $K>1/2$ and tilt-angle $\left|\theta\right|<\pi/2$,
Eq.~\eqref{eqn:scaling1} implies that the topological phase can always
be restored with a supercurrent such that $|\alpha-\delta k_{F}|<\delta p_{c}$.

We consider now the experimental consequences of our findings in tunneling transport spectroscopy \cite{Law,Flensberg,Wimmer,Liu}.
We consider a single normal metallic lead with a bias-voltage $eV_{b}$,
weakly coupled to the left end of the NW via the tunneling Hamiltonian
$H_{T}=\sum_{p,q}'T_{p,q}d_{p}^{\dag}\hat{c}_{q}+\sum_{p,j}T_{p,j}d_{p}^{\dag}\gamma_{j}+\text{H.c.}$,
where $T_{\mu\nu}$ are the tunneling coefficients, $d_{p}$ is the electron
annihilation operator in metallic lead, $\gamma_{j}$ are the second-quantization
MZBS operators localized at the left ($j=L$) and right ($j=R$) ends
of the NW. The sum $\sum'$ runs over the 1D-bulk states in the NW,
and the coupling coefficients $|T_{p,L}|\gg|T_{p,R}|$ due to the
exponentially localized Majorana wave functions. In the topological
phase, both the MZBS and the 1D-bulk continuum modes in the NW contribute
to the tunnel current $I$. Using the Keldysh formalism, we obtain
the tunnel current from $I=-e\dot{N}=-\frac{ie}{\hbar}\left[H_{T},N\right]$,
where $N=\sum_{p}d_{p}^{\dag}d_{p}$ is the number of electrons in
the metallic lead. Following Refs. \cite{Flensberg,Liu} we obtain
the expression
\begin{eqnarray}
I & = & \frac{e^{2}}{h}\int d\omega\mbox{Tr}\left[\Gamma^{e}{\cal G}^{R}\left(\omega\right)\Gamma^{h}{\cal G}^{A}\left(\omega\right)\right]\left[1-f\left(\omega-eV_b\right)\right]\nonumber \\
 &  & +\frac{e^{2}}{h}\int d\omega\Gamma\left(\omega\right)N\left(\omega\right)\left[1-f\left(\omega-eV_b\right)\right],\label{eqn:current1}
\end{eqnarray}
where $f\left(\omega\right)$ is Fermi distribution function and
the trace is taken in the subspace spanned by $\gamma_{j}$
modes. The retarded and advanced Majorana Green's functions ${\cal G}^{R}\left(\omega\right)=\left[{\cal G}^{A}\left(\omega\right)\right]^{\dag}$
and $\left[{\cal G}^{R}\left(\omega\right)\right]^{-1}=\omega/2+i\left[\Gamma^{e}\left(\omega\right)+\Gamma^{h}\left(\omega\right)\right]/2$,
where $\Gamma_{ij}^{e}\left(\omega\right)=\Gamma_{ij}^{h*}\left(-\omega\right)=2\pi\sum_{p}T_{p,i}T_{p,j}^{*}\delta\left(\omega-\varepsilon_{p}\right)$
are the self-energies, and $\varepsilon_{p}$ the single-electron
dispersion relation in the metallic lead. The second term in the right
hand side of Eq. \eqref{eqn:current1} represents the contribution
from 1D-bulk states, where $\Gamma\left(\omega\right)=2\pi\sum_{p}|T_{p,q}|^{2}\delta\left(\omega-\varepsilon_{p}\right)$,
and $N\left(\omega\right)$ is the 1D -bulk density of states in the
NW.

\begin{figure}[ht]
\includegraphics[width=0.8\columnwidth]{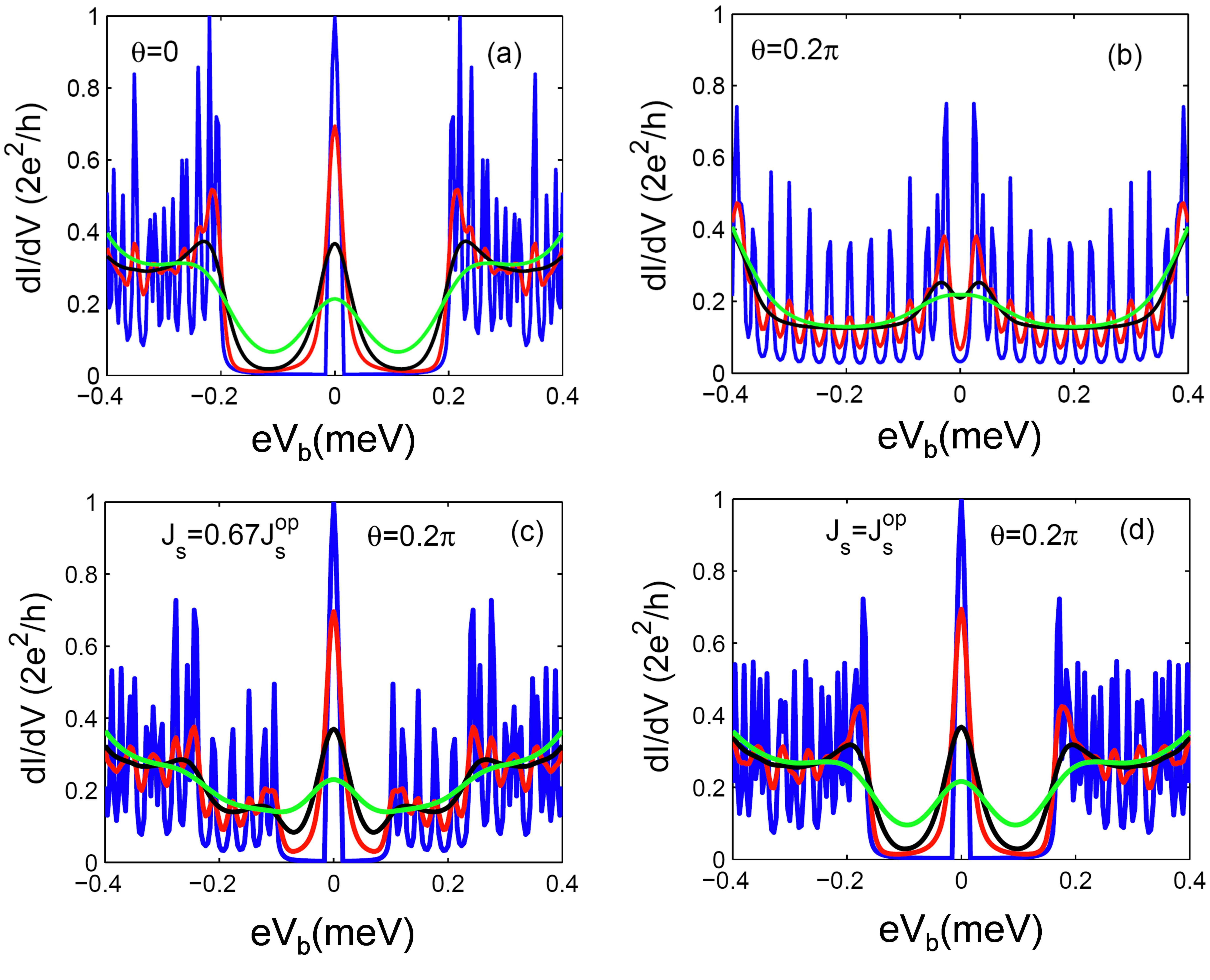} \caption{(Color online) $dI/dV$ for (a) $\theta=0$ and (b) $\theta=0.2\pi$ with
$J_{s}=0$. (c,d) Restoring the ZBP at $\theta=0.2\pi$
by supercurrents. The blue, red, black, and green curves correspond
to the temperature $T=0$, $60$mK, $180$mK, and $360$mK, respectively.
Other parameters are $V_{0}=1.0$meV, $E_{{\rm so}}=0.1$meV,
$\Delta_{s}=0.5$meV, and the tunneling energies $|\Gamma_{LL}^{e,h}|\sim|\Gamma|=0.005$meV.}
\label{DTC}
\end{figure}

Numerical results of $dI/dV$ are plotted in Fig.~\ref{DTC} (a-d)
at different temperatures. For $J_{s}=0$, a ZBP is obtained
when $\theta<\theta_{c}\approx0.168\pi$ [Fig.~\ref{DTC}
(a)], and disappears when $\theta>\theta_{c}$ [Fig.~\ref{DTC} (b)].
This result is consistent with the experimental observation in Ref. \cite{Kouwenhoven}. Fig.~\ref{DTC} (c,d) shows
that the Majorana-carrying phase is restored by a finite supercurrent
along $+x$ direction at $\theta=0.2\pi$, and maximizes at $J_{s}=J_{s}^{{\rm op}}\approx0.039J_{c}$
with $\xi\sim10$nm (Fig.~\ref{DTC} (d)) \cite{McCambridge}. The
ZBP in the $dI/dV$ spectra is clearly restored, indicating the reemergence
of MZBS after the bulk SC gap reopens. Further increasing $J_{s}$
again reduces the bulk gap (refer to Fig.~\ref{fig1} (b)). We confirm that the ZBP is $2e^{2}/h$ at $T=0$,
when the tunneling coefficients are small relative to the superconducting
bulk gap, and is strongly suppressed by thermal broadening. The disappearance and restoration of the ZBP provide useful experimental tests for topological
superconductivity in the lab.

\begin{figure}[t]
\includegraphics[width=0.8\columnwidth]{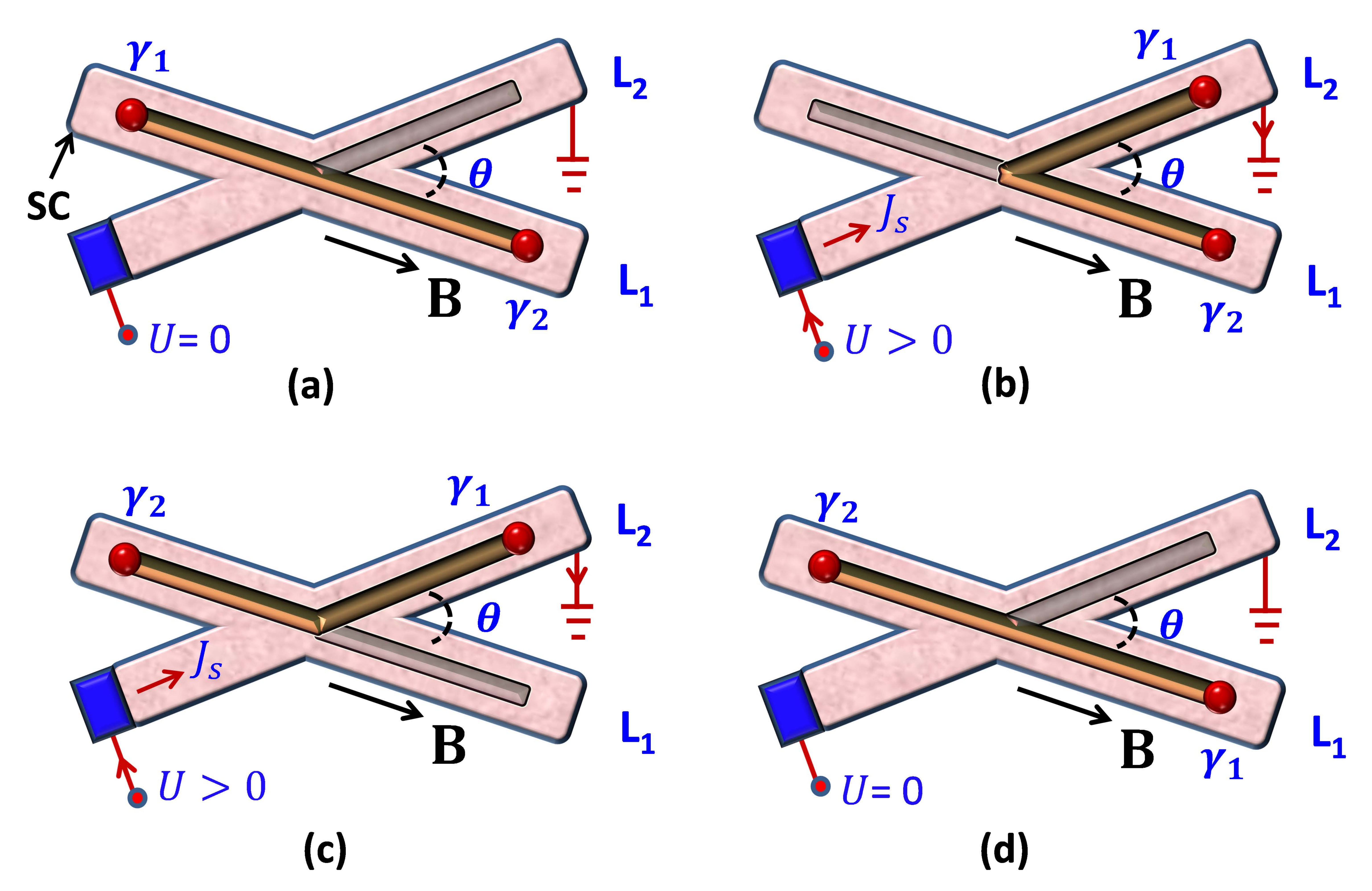} \caption{(Color online) Supercurrent assisted braiding (SAB) of two MZBS
in a ``\textbf{y}'' junction. (a) For $\theta>\theta_{c}$, the
NW segment $L_{2}$ is initially in the trivial phase at $J_{s}=0$.
(b) Move the MZBS $\gamma_{1}$ to NW $L_{2}$ by applying a supercurrent
$J_{s}=J_{s}^{{\rm op}}$ in $L_{2}$. (c) Move $\gamma_{2}$
to the original position of $\gamma_{1}$. (d) Move $\gamma_{1}$
to the NW $L_{1}$, and then turn off the supercurrent.}
\label{network}
\end{figure}

Finally we propose an important application of our findings to the
braiding of MZBS, as needed in TQC. For a 1D system, the braiding
operation of MZBS in a single NW is not well defined, and the minimum
requirement to exchange two MZBS is to consider a ``T''
or ``Y'' junction composed of several NW segments \cite{Alicea1,Clarke}. 
A realistic 2D/3D network of MZBS applicable for TQC can in principle
be constructed by putting together multiple NW junctions \cite{Halperin}.
However, in such a network  some of the NW segments are unavoidably
misaligned with the external magnetic field, therefore breaking the SIS in those NWs. Thus, being
able to drive all NWs deep into topological phase then becomes questionable,
bringing an inevitable difficulty to braid MZBS.
To resolve this problem, we introduce the SAB scheme, shown in Fig.~\ref{network} (a-d) for a ``\textbf{y}''-junction.
Here the spin quantization axis of a Rashba
SO coupling is perpendicular to the NW and parallel to
the SC plane (interface of the SC/NW heterostructure) \cite{Kouwenhoven}.
To minimize orbital effects, the external magnetic
field $\bold B$ must lie in the SC plane \cite{note1}, therefore
breaking SIS for at least one of the two NW segments.
If the $\bold B$ field is applied along the NW segment $L_1$ (Fig.~\ref{network}), the segment $L_2$ is
topologically trivial at $J_s=0$ when $\theta>\theta_c$.
On the other hand, to avoid the existence of low energy excitations at the
intersection of $L_{1,2}$, the tilt angle $\theta$ must be as close to $\pi/2$ as possible \cite{Alicea1}.
For the same parameters as in Fig.~\ref{fig1},
the critical angle is $\theta_{c}\approx0.168\pi$ (cf. also Ref. \cite{note0}).
Then for $\theta=0.2\pi$, the NW $L_2$ is already in the trivial phase without
applying a supercurrent [Fig.~\ref{network}(a)], and we next exchange two MZBS $\gamma_{1,2}$
localized on the ends of $L_{1}$. To perform the braiding of $\gamma_{1,2}$,
we apply a $J_{s}=J_{s}^{{\rm op}}\approx0.039J_{c}$
along $L_{2}$ (with  $\xi\sim10$nm for NbTi  \cite{McCambridge}) and move adiabatically first $\gamma_{1}$ to NW $L_{2}$ by gate control [Fig.~\ref{network}
(b)]. Then we move $\gamma_{2}$ to the original position of $\gamma_{1}$
[Fig.~\ref{network} (c)]. Finally $\gamma_{1}$ is shuttled to $L_{1}$,
completing exchange with $\gamma_{2}$, and the supercurrent is turned
off after braiding [Fig.~\ref{network}(d)].
It is noteworthy that
supercurrent is needed only in the intermediate process of the
braiding operation.
Applying the SAB to generic 2D or 3D Majorana networks can provide vast flexibility for the realistic TQC with MZBS.

In summary, we have studied the disappearance and reemergence of MZBS
in Majorana quantum wires with broken SIS, under the simultaneous effects of
a tilted magnetic field and supercurrents. We have shown the robustness
of these findings against the presence of e-e interactions, providing new insights
into the study of correlation effects in 1D topological SCs with broken SIS.
Finally, we introduced a supercurrent-assisted braiding of MZBS,
which has crucial applications to the realistic Majorana-fermion-based quantum computation.

\begin{acknowledgments}
We thank K. Zuo, L. Kouwenhoven, Patrick A. Lee, M. Cheng, K. T. Law, C. Wang and A. Iucci for helpful communications. We acknowledge support from  JQI-NSF-PFC, Microsoft-Q, and DARPA-QuEST.
\end{acknowledgments}
 \bibliographystyle{apsrev}


\newpage
\begin{widetext}

\begin{appendix}

\section{Supplementary Material for ``Manipulating Majorana Fermions in Quantum
Nanowires with Broken Inversion Symmetry"}

In this Supplementary Material we provide technical details on the bosonization method applied to the quantum nanowire with asymmetric dispersion relation (i.e., $\left|v_{R}\right|\neq\left|v_{L}\right|$), and give details on the derivation of the RG flow equations.
The methods used in this Supplemental Material are standard bosonization
and RG techniques that are explained in the usual textbooks \cite{apgiamarchi_book_1d,cardy_book_renormalization}

\subsection{Bosonization and diagonalization of the interacting system}

We start with the Hamiltonian for the lower subband $H=H_{\text{kin}}+H_{\text{int}}+H_{\text{p}},$
written in real space representation as

\begin{align*}
H_{\text{kin}} & =\int_{-L/2}^{L/2}dx\;\left[\left|v_{R}\right|:c_{R}^{\dagger}\left(x\right)\left(-i\partial_{x}\right)c_{R}\left(x\right):\right.\left.-\left|v_{L}\right|:c_{L}^{\dagger}\left(x\right)\left(-i\partial_{x}\right)c_{L}\left(x\right):\right],\\
H_{\text{int}} & =\pi U\int_{-L/2}^{L/2}dx\;:c_{R}^{\dagger}\left(x\right)c_{R}\left(x\right)::c_{L}^{\dagger}\left(x\right)c_{L}\left(x\right):,\\
H_{\text{p}} & =-i\Delta_{\text{p}}\int_{-L/2}^{L/2}dx\;\left[e^{i\alpha x}c_{R}\left(x\right)c_{L}\left(x\right)-\text{H.c.}\right],
\end{align*}
where $U$ represents a short-distance Coulomb interaction,
$\alpha$ is the phase-gradient induced by a supercurrent in the bulk
SC, and the notation $:\ \dots \ :$ means normal ordering. Here for definiteness we assume $v_{R}>0$ and $v_{L}<0$. We
now introduce the Abelian bosonization method \cite{apgiamarchi_book_1d},
and write down the fermionic annihilation field operators as
\begin{align*}
c_{R}\left(x\right) & =\frac{F_{R}}{\sqrt{2\pi a}}e^{ik_{R}x}e^{i\phi_{R}\left(x\right)},\\
c_{L}\left(x\right) & =\frac{F_{L}}{\sqrt{2\pi a}}e^{-ik_{L}x}e^{i\phi_{L}\left(x\right)},
\end{align*}
where $a$ is the short distance cutoff of the theory, $\phi_{R/L}\left(x\right)$
are chiral bosonic fields which obey the commutation relations $\left[\phi_{R/L}\left(x\right),\phi_{R/L}\left(y\right)\right]=\pm\frac{i\pi}{2}\text{sgn}\left(x-y\right)$,
and $F_{R/L}$ are Klein factors that obey $\left\{ F_{a},F_{b}\right\} =\left\{ F_{a}^{\dagger},F_{b}^{\dagger}\right\} =0$
and $\left\{ F_{a},F_{b}^{\dagger}\right\} =\delta_{ab}$ (with $\{a,b\}=\{R,L\}$),
and therefore ensure fermionic anticommutation relations of $c_{R/L}\left(x\right)$.
Note that due to the asymmetry in the spectrum, the Fermi momenta
$k_{R}\neq k_{L}$. In terms of the fields $\phi_{R/L}\left(x\right)$
the Hamiltonian reads
\begin{align*}
H_{\text{kin}} & =\frac{1}{4\pi}\int_{-L/2}^{L/2}dx\;\left[\left|v_{R}\right|\left(\partial_{x}\phi_{R}\right)^{2}+\left|v_{L}\right|\left(\partial_{x}\phi_{L}\right)^{2}\right]\\
H_{\text{int}} & =\frac{U}{4\pi}\int_{-L/2}^{L/2}dx\;\left(\partial_{x}\phi_{R}\right)\left(-\partial_{x}\phi_{L}\right),\\
H_{\text{p}} & =-\frac{i\Delta_{\text{p}}}{2\pi a}\int_{-L/2}^{L/2}dx\;\left[e^{i\delta px}e^{i\left(\phi_{R}+\phi_{L}\right)}-\text{H.c.}\right],
\end{align*}
where $\delta p\equiv\alpha+k_{R}-k_{L}$, and where we have dropped
the Klein factors since they are not relevant in what follows. In
the absence of pairing (i.e., $\Delta_{p}=0$), the Hamiltonian $H_{0}=H_{\text{kin}}+H_{\text{int}}$
is quadratic in $\phi_{R/L}$, and therefore it can be diagonalized by
solving the equation\cite{wen_1,apFernandez02_Asymmetric_LL}

\begin{align}
\det\left[\begin{array}{cc}
\left|v_{R}\right|-\frac{v_{\pm}}{\text{sgn}\left(v_{R}\right)} & \frac{U}{2}\\
\frac{U}{2} & \left|v_{L}\right|-\frac{v_{\pm}}{\text{sgn}\left(v_{L}\right)}
\end{array}\right] & =0.\label{eq:det}
\end{align}
This problem is analogous to the interacting edge modes of the fractional
quantum Hall effect (FQHE) at filling factor $2/3$, where a chiral mode
with filling 1 interacts with a counterpropagating mode with filling
$-1/3$ \cite{wen_1}. The solutions of (\ref{eq:det}) are given by
two new counterpropagating modes with velocities\cite{wen_1,apFernandez02_Asymmetric_LL}
\begin{align}
v_{\pm} & =\frac{1}{2}\left[\left|v_{R}\right|-\left|v_{L}\right|\pm\sqrt{\left(\left|v_{R}\right|+\left|v_{L}\right|\right)^{2}-U^{2}}\right],\nonumber \\
 & =v\left[\eta\pm\sqrt{1-g^{2}}\right],\label{eq:v_pm}
\end{align}
where we have defined the average velocity $v=\left(\left|v_{R}\right|+\left|v_{L}\right|\right)/2$
, the asymmetry parameter $\eta=\left(\left|v_{R}\right|-\left|v_{L}\right|\right)/\left(\left|v_{R}\right|+\left|v_{L}\right|\right)$
and the interaction parameter $g=U/\left(\left|v_{R}\right|+\left|v_{L}\right|\right)=2U/v$.
From Ref. \onlinecite{apFernandez02_Asymmetric_LL}, we know that the
regime of stability of the Luttinger liquid is $\eta^{2}+g^{2}<1$.
The new eigenmodes are given by
\begin{align}
\left(\begin{array}{c}
\phi_{+}\\
\phi_{-}
\end{array}\right) & =\left(\begin{array}{cc}
\cosh\chi & \sinh\chi\\
\sinh\chi & \cosh\chi
\end{array}\right)\left(\begin{array}{c}
\phi_{R}\\
-\phi_{L}
\end{array}\right),\label{eq:eigenmodes}
\end{align}
where the parameter $\chi$ is defined through $\tanh\chi=g^{-1}\left[1-\sqrt{1-g^{2}}\right].$
In terms of $\phi_{\pm}$, the Hamiltonian $H_{0}$ writes
\begin{align}
H_{0} & =\frac{1}{4\pi}\int_{-L/2}^{L/2}dx\;\left[\left|v_{+}\right|\left(\partial_{x}\phi_{+}\right)^{2}+\left|v_{-}\right|\left(\partial_{x}\phi_{-}\right)^{2}\right].\label{eq:H_bosonic_diagonal}
\end{align}
Note that the new fields $\phi_{\pm}$ obey the usual commutation
relations for chiral fields, $\left[\phi_{\pm}\left(x\right),\phi_{\pm}\left(y\right)\right]=\pm\frac{i\pi}{2}\text{sgn}\left(x-y\right)$.
From here we see that the interacting system is still described by
a Tomonaga Luttinger liquid (TLL) model with asymmetric dispersion
relation, and consequently there are two branches of 1D plasmon excitations,
traveling with different velocities $v_{+}$ and $v_{-}$ \cite{wen_1,apFernandez02_Asymmetric_LL}.

To make contact with the standard notation in terms of non-chiral
fields $\vartheta,\varphi$ (as in the main manuscript), we introduce
the change of variables

\begin{align}
\phi_{R/L} & =\mp\varphi+\vartheta.\label{eq:non-chiral}
\end{align}
From Eqs. (\ref{eq:eigenmodes}) and (\ref{eq:non-chiral}), we obtain
the relation
\begin{align}
 & \left(\begin{array}{c}
\varphi\\
\vartheta
\end{array}\right)=\left(\begin{array}{cc}
\frac{-\cosh\chi+\sinh\chi}{2} & \frac{-\cosh\chi+\sinh\chi}{2}\\
\frac{\cosh\chi+\sinh\chi}{2} & \frac{-\cosh\chi-\sinh\chi}{2}
\end{array}\right)\left(\begin{array}{c}
\phi_{+}\\
\phi_{-}
\end{array}\right).\label{eq:change_of_basis}
\end{align}
We can now rewrite Eq. (\ref{eq:H_bosonic_diagonal}) in terms of
the fields $\varphi,\vartheta$ as%
\begin{align}
H_{0} & =\frac{v}{2\pi}\int_{-L/2}^{L/2}dx\;\left[\frac{\left(\partial_{x}\varphi\right)^{2}}{K}+K\left(\partial_{x}\vartheta\right)^{2}-2\eta\partial_{x}\varphi\partial_{x}\vartheta\right],\label{eq:H0_non_chiral}
\end{align}
where $K=\frac{\cosh\chi-\sinh\chi}{\cosh\chi+\sinh\chi}=\frac{\sqrt{1-g}}{\sqrt{1+g}}$.

\subsection{Derivation of the RG equations in the presence of pairing}

We now focus on the effect of the superconducting term $H_{\text{p}}$,
and study the limit when $H_{\text{p}}$ is a perturbation to the
fixed point Hamiltonian $H_{0}$. We start by writing the total partition
function of the system

\begin{align}
Z & =\text{Tr }e^{-\left(H_{0}+H_{\text{p}}\right)/T}=\int\prod_{\nu=\pm}\mathcal{D}\left[\phi_{\nu}\right]\: e^{-S_{0}-S_{\text{p}}},\label{eq:Z}
\end{align}
where $S_{0}$ is the Euclidean action corresponding to Hamiltonian
$H_{0}$ Eq. (\ref{eq:H_bosonic_diagonal})

\begin{align}
S_{0} & =\frac{1}{4\pi}\sum_{\nu=\pm}\int_{-L/2}^{L/2}dx\int_{-\beta/2}^{\beta/2}d\tau\;\partial_{x}\phi_{\nu}\left(x,\tau\right)\left[-\left(\nu\right)i\partial_{\tau}\phi_{\nu}\left(x,\tau\right)+\left|v_{\nu}\right|\partial_{x}\phi_{\nu}\left(x,\tau\right)\right],\label{eq:S0}
\end{align}
where $\tau$ is the imaginary time, and $\beta=1/T$ is the inverse
temperature. In the following, we focus in the limit $L\rightarrow\infty$
and $T\rightarrow0$. The term $S_{\text{p}}$ is the pairing interaction
\begin{align}
S_{\text{p}} & =\frac{y}{2\pi i}\int\frac{d^{2}\mathbf{r}}{a^{2-1/K}}\;\left[e^{i\delta px}V_{+}\left(\mathbf{r}\right)V_{-}^{*}\left(\mathbf{r}\right)-\text{H.c.}\right],\label{eq:Sp}
\end{align}
where we have introduced the vertex operator $V_{\pm}\left(x,\tau\right)\equiv a^{-1/2K}\exp\left[i\phi_{\nu}\left(x,\tau\right)/\sqrt{K}\right]$
and the dimensionless pairing parameter $y=\Delta_{\text{p}}a/v$. Note that in (\ref{eq:Sp}) we have also introduced the compact
notation $\mathbf{r}=\left(x,v\tau\right)$.

We now return to Eq. (\ref{eq:Z}) and expand the partition function
up to second order in powers of $y$
\begin{align}
Z & =Z_{0}\times\left\{ 1+\frac{1}{2!}\left(\frac{y}{2\pi}\right)^{2}\int_{\left|\mathbf{r}_{1}-\mathbf{r}_{2}\right|>a}\frac{d^{2}\mathbf{r}_{1}d^{2}\mathbf{r}_{2}}{a^{4-2/K}}\;\right. \left.\times\left[e^{i\delta p\left(x_{1}-x_{2}\right)}\prod_{\nu=\pm}\left\langle V_{\nu}\left(\mathbf{r}_{1}\right)V_{\nu}^{*}\left(\mathbf{r}_{2}\right)\right\rangle _{0}+\text{H.c.}\right]\right\} \label{eq:Z_2nd_order}
\end{align}
 where the averages are taken with respect to the fixed-point action
$S_{0}$, and where we have used that $\left\langle V_{\nu}\left(\mathbf{r}\right)\right\rangle _{0}=0$.
The correlators are $\left\langle V_{\nu}\left(\mathbf{r}_{1}\right)V_{\nu^{\prime}}^{*}\left(\mathbf{r}_{2}\right)\right\rangle _{0}=\left[\left(\left|x\right|+a\right)^{2}+\left(\left|v_{\nu}\right|\tau\right)^{2}\right]^{-1/2K}$
for $\nu=\nu^{\prime}$, and zero otherwise \cite{apgiamarchi_book_1d}.

We now implement the RG transformation by performing an infinitesimal
change in the microscopic cutoff $a$, and asking how the couplings
$\left\{ K,v,\eta,y\right\} $ of the model should change in order
to preserve the partition function $Z$. It is convenient to parametrize
the RG transformation with a dimensionless continuous variable $\ell$,
i.e., $a=a\left(\ell\right)\equiv a_{0}e^{\ell}$. In this way, the
couplings of the model become functions of $\ell$ through their dependence
on $a\left(\ell\right)$: $\left\{ K,v,\eta,y\right\} \rightarrow\left\{ K\left(\ell\right),v\left(\ell\right),\eta\left(\ell\right),y\left(\ell\right)\right\} $.
We now focus on the infinitesimal transformation $a\left(\ell\right)\rightarrow a\left(\ell+d\ell\right)\simeq a\left(\ell\right)\left[1+d\ell\right]$,
and demand that the equation

\begin{align}
Z\left(\ell\right) & =Z\left(\ell+d\ell\right),\label{eq:Z_RG_transf}
\end{align}
is satisfied \cite{cardy_book_renormalization,apgiamarchi_book_1d}.
To simplify the notation, we denote the integral over $\mathbf{r}_{1}$
and $\mathbf{r}_{2}$ in (\ref{eq:Z_2nd_order}) as

\begin{align}
\left\langle I\left(\ell\right)\right\rangle _{0} & =\frac{y^{2}\left(\ell\right)}{8\pi^{2}}\int_{\left|\mathbf{r}_{1}-\mathbf{r}_{2}\right|>a\left(\ell\right)}\frac{d^{2}\mathbf{r}_{1}d^{2}\mathbf{r}_{2}}{a^{4-2/K}\left(\ell\right)}\left[e^{i\delta p\left(x_{1}-x_{2}\right)}\right. \left.\times\prod_{\nu=\pm}\left\langle V_{\nu}\left(\mathbf{r}_{1}\right)V_{\nu}^{*}\left(\mathbf{r}_{2}\right)\right\rangle _{0}+\text{H.c.}\right].\label{eq:I}
\end{align}
In terms of $\left\langle I\left(\ell\right)\right\rangle _{0}$,
Eq. (\ref{eq:Z_RG_transf}) writes $Z_{0}\left(\ell\right)\left[1+I\left(\ell\right)\right]=Z_{0}\left(\ell+d\ell\right)\left[1+I\left(\ell+d\ell\right)\right].$
We now split $I\left(\ell+d\ell\right)$ into
\begin{align}
\left\langle I\left(\ell+d\ell\right)\right\rangle _{0} & =y^{2}\left(\ell+d\ell\right)\times\left[\int_{\left|\mathbf{r}_{1}-\mathbf{r}_{2}\right|>a\left(\ell\right)}\right. \left.-\int_{a\left(\ell\right)\left[1+d\ell\right]>\left|\mathbf{r}_{1}-\mathbf{r}_{2}\right|>a\left(\ell\right)}\right],\label{eq:I_ldl}
\end{align}
where we have made explicit the dependence on $y\left(\ell\right)$.
Note that the first term in the r.h.s. gives back $\left\langle I\left(\ell\right)\right\rangle _{0}$,
provided we perform the change $y\left(\ell+d\ell\right)=y\left(\ell\right)e^{\left(2-1/K\right)d\ell}$.
On the other hand, the second term in the r.h.s. in (\ref{eq:I_ldl})
can be written 
\begin{align}
\left\langle I_{2}\left(\ell+d\ell\right)\right\rangle _{0} & =-\frac{y^{2}\left(\ell\right)e^{\left(4-2/K\right)d\ell}}{8\pi^{2}}\int d^{2}\mathbf{R}\int_{a\left(\ell+d\ell\right)>r>a\left(\ell\right)}\frac{d^{2}\mathbf{r}\; e^{i\delta px}}{a^{4-2/K}\left(\ell+d\ell\right)}\prod_{\nu=\pm}\left\langle V_{\nu}\left(\mathbf{R}+\frac{\mathbf{r}}{2}\right)V_{\nu}^{*}\left(\mathbf{R}-\frac{\mathbf{r}}{2}\right)\right\rangle _{0}+\text{H.c.},\label{eq:I2}
\end{align}
where we have introduced relative and center-of-mass coordinates,
$\mathbf{r}=\mathbf{r}_{1}-\mathbf{r}_{2}$ and $\mathbf{R}=\frac{1}{2}\left(\mathbf{r}_{1}+\mathbf{r}_{2}\right)$.
This term renormalizes the fixed point action $S_{0}\left(\ell+d\ell\right)$.
To see this, we first need to extract the operator content of $V_{\nu}\left(\mathbf{R}+\frac{\mathbf{r}}{2}\right)V_{\nu}^{*}\left(\mathbf{R}-\frac{\mathbf{r}}{2}\right)$
in the limit $\mathbf{r}\rightarrow0$, and to that end we perform
the operator product expansion (OPE) \cite{cardy_book_renormalization}:
$V_{\nu}\left(\mathbf{R}+\frac{\mathbf{r}}{2}\right)V_{\nu}^{*}\left(\mathbf{R}-\frac{\mathbf{r}}{2}\right)\xrightarrow[\mathbf{r}\rightarrow0]{}a^{-1/K}\left(\ell\right)\sum_{n=0}^{\infty}\frac{1}{n!}\left[\frac{i\left(x-i\tau v_{\nu}\right)}{\sqrt{K}}\partial_{x}\phi_{\nu}\left(\mathbf{R}\right)\right]^{n},$
where we have used the equation of motion for chiral fields $\partial_{\tau}\phi_{\nu}\left(x,\tau\right)=-iv_{\nu}\partial_{x}\phi_{\nu}\left(x,\tau\right)$,
obtained from minimization of $S_{0}$ in (\ref{eq:S0}). It is now convenient to rewrite (\ref{eq:I2}) in terms of the non-chiral
fields $\left(\varphi,\vartheta\right)$ using (\ref{eq:change_of_basis})
and expressing the integral over $\mathbf{r}$ in cylindrical coordinates
$x=r\cos\Theta$, $y=r\sin\Theta$. 
At first order in $d\ell$,
we obtain 
\begin{align*}
I_{2}\left(\ell+d\ell\right) & =-\frac{y^{2}\left(\ell\right)d\ell}{2\pi^{2}}\int d^{2}\mathbf{R}\times\left[\frac{\left(\partial_{x}\varphi\left(\mathbf{R}\right)\right)^{2}}{K^{2}\left(\ell\right)}\int_{0}^{2\pi}d\Theta e^{i\delta pr\cos\Theta}\sin^{2}\Theta-\left(\partial_{x}\vartheta\left(\mathbf{R}\right)\right)^{2}\right.\\
 & \left.\times\int_{0}^{2\pi}d\Theta e^{i\delta p\cos\Theta}\left(\cos^{2}\Theta-\eta^{2}\sin^{2}\Theta\right)-\frac{4\eta\left(\ell\right)}{1+K^{2}\left(\ell\right)}\partial_{x}\varphi\left(\mathbf{R}\right)\partial_{x}\vartheta\left(\mathbf{R}\right)\int_{0}^{2\pi}d\Theta e^{i\delta p\cos\Theta}\sin^{2}\Theta\right],
\end{align*}
where we have approximated $r\simeq a\left(\ell\right)$. Performing
the angular integral yields
\begin{align*}
I_{2}\left(\ell+d\ell\right) & =-\frac{y^{2}\left(\ell\right)d\ell}{2\pi}\int d^{2}\mathbf{R}\times\left\{ \frac{\left(\partial_{x}\varphi\left(\mathbf{R}\right)\right)^{2}}{K^{2}\left(\ell\right)}\left[J_{0}\left(\delta pa\left(\ell\right)\right)+J_{2}\left(\delta pa\left(\ell\right)\right)\right]-\left(\partial_{x}\vartheta\left(\mathbf{R}\right)\right)^{2}\left[\left(1-\eta^{2}\left(\ell\right)\right)J_{0}\left(\delta pa\left(\ell\right)\right)\right.\right.\\
 & \left.\left.-\left(1+\eta^{2}\left(\ell\right)\right)J_{2}\left(\delta pa\left(\ell\right)\right)\right]-\frac{4\eta\left(\ell\right)}{1+K^{2}\left(\ell\right)}\partial_{x}\varphi\left(\mathbf{R}\right)\partial_{x}\vartheta\left(\mathbf{R}\right)\left[J_{0}\left(\delta pa\left(\ell\right)\right)+J_{2}\left(\delta pa\left(\ell\right)\right)\right]\right\} .
\end{align*}
Reexponentiating this term in Eq. (\ref{eq:Z_2nd_order}) and returning
to Eq. (\ref{eq:Z_RG_transf}) yields
\begin{align}
\int\mathcal{D}\left[\varphi,\vartheta\right]e^{-S_{0}\left(\ell\right)}\left[1+I\left(\ell\right)\right] & =\int\mathcal{D}\left[\varphi,\vartheta\right]e^{-S_{0}\left(\ell+d\ell\right)+I_{2}\left(\ell+d\ell\right)}\left[1+I\left(\ell\right)\right].\label{eq:RG_Z}
\end{align}
This equation is satisfied imposing $S_{0}\left(\ell\right)=S_{0}\left(\ell+d\ell\right)-I_{2}\left(\ell+d\ell\right)$. Using the relation $y\left(\ell+d\ell\right)=y\left(\ell\right)e^{\left(2-1/K\right)d\ell}$
and matching the coefficients of the terms $\left(\partial_{x}\varphi\right)^{2}$,
$\left(\partial_{x}\vartheta\right)^{2}$ and $\partial_{x}\varphi\partial_{x}\vartheta$
in (\ref{eq:H0_non_chiral}) result in the RG flow equations 
\begin{align}
\frac{dy}{d\ell} & =\left[2-K^{-1}\right]y\left(\ell\right).\label{eq:RG_y}\\
\frac{dK}{d\ell} & =\frac{y^{2}\left(\ell\right)}{2}\left[\left(2-\eta^{2}\left(\ell\right)\right)J_{0}\left(\delta pa\left(\ell\right)\right)-\eta^{2}\left(\ell\right)J_{2}\left(\delta pa\left(\ell\right)\right)\right]\label{eq:RG_K}\\
\frac{dv}{d\ell} & =-\frac{y^{2}\left(\ell\right)v\left(\ell\right)}{2K\left(\ell\right)}\left[\eta^{2}\left(\ell\right)J_{0}\left(\delta pa\left(\ell\right)\right)+\left(2+\eta^{2}\left(\ell\right)\right)J_{2}\left(\delta pa\left(\ell\right)\right)\right]\label{eq:RG_u}\\
\frac{d\eta}{d\ell} & =y^{2}\left(\ell\right)\eta\left(\ell\right)\left[\left(\frac{2}{1+K^{2}\left(\ell\right)}+\frac{\eta^{2}\left(\ell\right)}{2K\left(\ell\right)}\right)J_{0}\left(\delta pa\left(\ell\right)\right)+\left(\frac{2}{1+K^{2}\left(\ell\right)}+\frac{2+\eta^{2}\left(\ell\right)}{2K\left(\ell\right)}\right)J_{2}\left(\delta pa\left(\ell\right)\right)\right].\label{eq:RG_eta}
\end{align}
Note that these RG equations are only perturbative
in $y\left(\ell\right)$, and are exact in $\eta\left(\ell\right)$.
We note that at the leading order the RG equation (\ref{eq:RG_y}) for $y\left(\ell\right)$ is independent of $\eta\left(\ell\right)$.
On the other hand, in the limit $\delta p a(\ell) \ll 1$, the RG equation \ref{eq:RG_eta} implies that the amplitude of $\eta(\ell)$ grows upon renormalization. Physically, this means that the band asymmetry is more important at lower energy scale. From Eq. (18) one finds that this effect can slow down the growth of $K\left(\ell\right)$, and therefore can be detrimental on the p-wave SC phase.
However, note that in the limit of small band-asymmetry $\eta(\ell)\ll1$, its effects become higher order processes $\sim\mathcal{O}\left(\eta^{2}y^{2}\right)$ in Eqs. (18) and (19), where the dominant order is $\mathcal{O}\left(y^{2}\right)$. Then a small $\eta$-term can only lead to minor quantitative corrections to the RG flows of $K$ and $u$, and do not affect the main results described in the manuscript.  For the typical parameter regime used in Fig. 1 of the main manuscript, one can verify that $\eta<0.01$ at all tilt angles. We therefore can safely neglect terms $\mathcal{O}\left(y^{2}\eta^{2}\right)$ in Eqs. (\ref{eq:RG_K}) and (\ref{eq:RG_u}), and approximate Eq. (\ref{eq:RG_eta}) by $d\eta/d\ell\approx0$. In this case,
all the dependence on $\eta\left(\ell\right)$ drops from the RG equations at leading order, and we recover the expressions in the main manuscript.

\bibliographystyle{apsrev}

\begin{thebibliography}{44}
\bibitem{note0} The typical magnitude of $\theta_c$ may depend on materials. For InAs wire in proximity to the SC material Al and for parameters $\Delta_s\sim40\mu$eV, $V_0\sim62.5\mu$eV, and $\mu\sim0$ (cf. Ref. \cite{Das}), it follows that $\theta_c\sim0.22\pi$.
\bibitem{note1} The orbital effects can generically harm the TQC. For example, for the type-II SC material NbTi to drive InSb nanowire deep into topological phase requires $B>0.2$T \cite{Kouwenhoven}, which is much larger than the lower critical field of NbTi (typically in the order of $10$mT). This effect complicates phase distribution of the induced SC order parameter in the Majorana network due to the formation of vortices in NbTi, and can lead to uncontrollable low energy excitations in the network. For type-I SC, e.g. Al applying the $\bold B$-field out-of-plane may even destroy the SC phase since the critical field is typically very low.
\end{thebibliography}

\end{appendix}

\end{widetext}


\end{document}